\documentclass[conference]{IEEEtran}
\usepackage{cite}

\usepackage{graphicx}
\usepackage{subfigure}
\usepackage{epstopdf}
\usepackage{upgreek}
\usepackage{amsmath}
\usepackage{amsthm}
\usepackage{marvosym}
\usepackage{amssymb}
\usepackage{bbm}
\usepackage{amsfonts,amssymb}
\usepackage{multirow}
\usepackage{enumitem}
\setlist{noitemsep, topsep=0em, leftmargin=*, labelsep=0.6em}
\usepackage{color}
\usepackage{algorithm}
\usepackage{algpseudocode}
\graphicspath{{Fig/}}
\theoremstyle{plain}

\usepackage{booktabs}
\usepackage{threeparttable}
\usepackage{makecell}
\usepackage{tabularx}

\usepackage{balance}
\usepackage{color,soul}

\usepackage[hyphens]{url}
\usepackage{hyperref}
\usepackage[hyphenbreaks]{breakurl}

{}
{}
{}

\begin{document}

\title{Toward 6G TK$\boldsymbol\upmu$ Extreme Connectivity: Architecture, Key Technologies and Experiments}

\author{\IEEEauthorblockN{
Xiaohu~You\IEEEauthorrefmark{2}\IEEEauthorrefmark{3}\IEEEauthorrefmark{4}$^*$,
Yongming~Huang\IEEEauthorrefmark{2}\IEEEauthorrefmark{3}$^*$,
Shengheng~Liu\IEEEauthorrefmark{2}\IEEEauthorrefmark{3},
Dongming~Wang\IEEEauthorrefmark{2}\IEEEauthorrefmark{3},
Junchao~Ma\IEEEauthorrefmark{2},
Chuan~Zhang\IEEEauthorrefmark{2}\IEEEauthorrefmark{3},\\
Hang~Zhan\IEEEauthorrefmark{2},
Cheng~Zhang\IEEEauthorrefmark{2}\IEEEauthorrefmark{3},
Jiao~Zhang\IEEEauthorrefmark{2},
Zening~Liu\IEEEauthorrefmark{2},
Jin~Li\IEEEauthorrefmark{2},
Min~Zhu\IEEEauthorrefmark{2},
Jianjie~You\IEEEauthorrefmark{2},
Dongjie~Liu\IEEEauthorrefmark{2},\\
Yang~Cao\IEEEauthorrefmark{2},
Shiwen~He\IEEEauthorrefmark{2},
Guanghui~He\IEEEauthorrefmark{2},
Fengyi~Yang\IEEEauthorrefmark{5},
Yang~Liu\IEEEauthorrefmark{5},
Jianjun~Wu\IEEEauthorrefmark{6},
Jianmin~Lu\IEEEauthorrefmark{6},
Ge~Li\IEEEauthorrefmark{4},\\
Xiaowu~Chen\IEEEauthorrefmark{4},
Wenguang~Chen\IEEEauthorrefmark{4},
Wen~Gao\IEEEauthorrefmark{4}}

\medskip

\IEEEauthorblockA{\IEEEauthorrefmark{2}Purple Mountain Laboratories, Nanjing 211111, China\\}
\IEEEauthorblockA{\IEEEauthorrefmark{3}National Mobile Communications Research Laboratory, Southeast University, Nanjing 210096, China\\}
\IEEEauthorblockA{\IEEEauthorrefmark{4}Peng Cheng Laboratory, Shenzhen 518066, China\\}
\IEEEauthorblockA{\IEEEauthorrefmark{5}China Telecom Research Institution, Beijing 102209, China\\}
\IEEEauthorblockA{\IEEEauthorrefmark{6}Huawei Technologies Company Ltd., Shenzhen 518066, China\\
*\Letter \; \{xhyu; huangym\}@seu.edu.cn}
}
	
\maketitle

\begin{abstract}
Sixth-generation (6G) networks are evolving towards new features and order-of-magnitude enhancement of systematic performance metrics compared to the current 5G. In particular, the 6G networks are expected to achieve extreme connectivity performance with ${\rm{Tbps}}$-scale data rate, ${\rm{Kbps/Hz}}$-scale spectral efficiency, and ${\rm{\upmu s}}$-scale latency. To this end, an original three-layer 6G network architecture is designed to realize uniform full-spectrum cell-free radio access and provide task-centric agile proximate support for diverse applications. The designed architecture is featured by super edge node (SEN) which integrates connectivity, computing, AI, data, etc. On this basis, a technological framework of pervasive multi-level (PML) AI is established in the centralized unit to enable task-centric near-real-time resource allocation and network automation. We then introduce a radio access network (RAN) architecture of full spectrum uniform cell-free networks, which is among the most attractive RAN candidates for 6G ${\rm{TK\upmu}}$ extreme connectivity. A few most promising key technologies, i.e., cell-free massive MIMO, photonics-assisted Terahertz wireless access and spatiotemporal two-dimensional channel coding are further discussed. A testbed is implemented and extensive trials are conducted to evaluate innovative technologies and methodologies. The proposed 6G network architecture and technological framework demonstrate exciting potentials for full-service and full-scenario applications.
\end{abstract}

\maketitle

\section{Introduction}
\label{sec1}

\begin{figure*}[!htb]
	\centering
	\includegraphics[width = 0.96\linewidth]{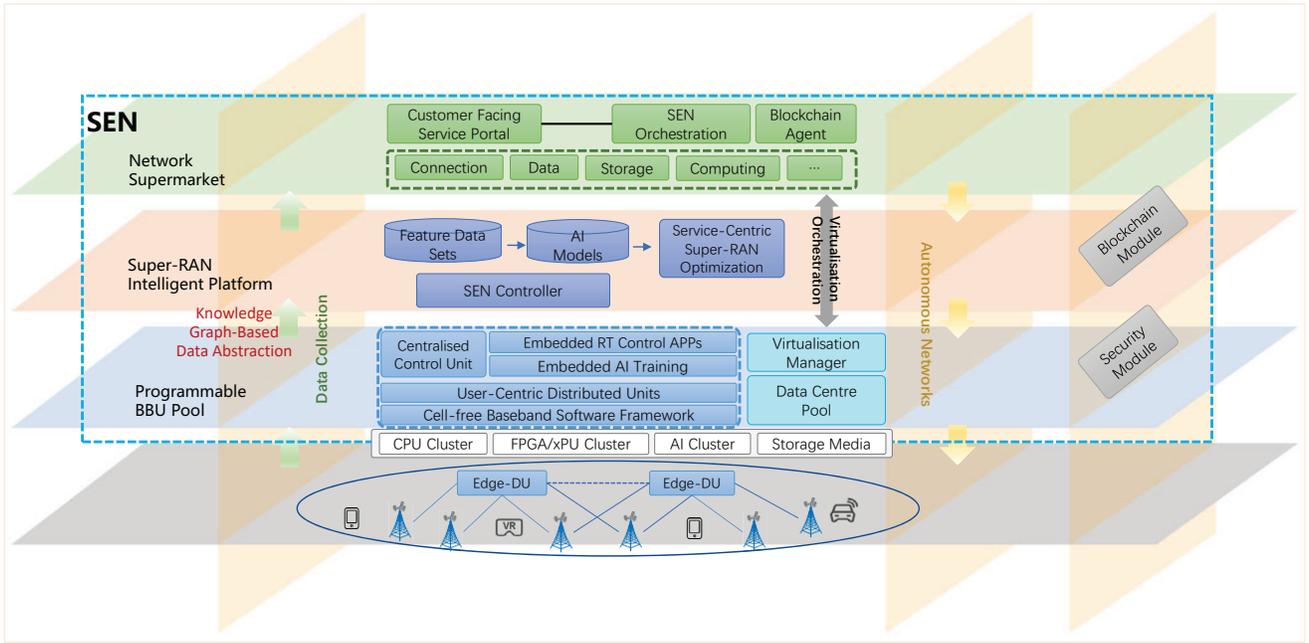}
    \vspace{-0.7em}
	\caption{SEN architecture with pBBUP.}
	\label{fig2_2}
\end{figure*}

While the roll-out of fifth-generation (5G) networks is speeding along at a spectacular pace, several sixth-generation (6G) research and development (R\&D) initiatives have been launched by the leading powers across the world such as the United States (US), European Union (EU), Korea, and China \cite{NSF2021, Hexa2020, Samsung2020}. Despite varied focuses, the major countries and institutions embrace the same visions, i.e., 6G will (i) further expand connectivity ability and service coverage on the basis of 5G; (ii) further strengthen its capability as the key enabler and deepen the penetration in diverse vertical applications; (iii) progress through the three stages --- establishment of overall vision and technology framework, standardization, and large-scale commercial deployment, respectively, around 2023, 2025, and 2030. 

The shared visions and expectations are built upon the increasingly stringent requirements in terms of network performance and intelligence imposed by the ongoing digitalization of society and economy. For instance, immerse extended reality (XR) in maintenance and repair situations requires ${\rm{Tbps}}$-scale data rate \cite{XR}; massive machine-type communications for Industry 4.0 requires ${\rm{Kbps/Hz}}$-scale spectral efficiency (SE) to guarantee quality of service \cite{5}; the expected latency and reliability for vehicular communications and industrial control are respectively at ${\rm{\upmu s}}$ level and $99.9999\%$ \cite{tariq2020speculative}. Aside from the key KPIs, game-changing features such as native AI for customized diverse applications and zero-touch network management are also envisioned in the future 6G. Nonetheless, all of the above is clearly beyond the capability of the current network architecture and technological framework.

In the light of the above observations, we coin a new acronym of three arguably the most important key performance indicators (KPIs) --- ${\rm{TK\upmu}}$ for 6G evolvement. Concretely, we deduce the following KPI estimates for the future 6G extreme connectivity. Specifically, 6G will require a (i) data rate of $1\;{\rm{Tbps}}$, (ii) SE of $1\;{\rm{Kbps/Hz}}$, and  (iii) ${\rm{\upmu s}}$-level latency for vast applications. In order to realise 6G ${\rm{TK\upmu}}$ extreme connectivity, holistic and systematic consideration of the hierarchical architecture as well as a comprehensive methodology and technological framework are essential \cite{11}.

In this context, the following efforts are contributed. First, we propose a novel task-centric three-layer architecture for 6G networks, which underpins the support provided by 6G for full-service and full-scenario applications. The proposed architecture features a decentralized model and employs super edge node (SEN) as the basic enabling facility to agilely allocate requested resources in the proximity of the mobile users. Together with the cell-free massive MIMO technologies deployed in the edge distributed unit (EDU), an extreme connectivity capability of ${\rm{TK\upmu}}$-level is yielded. 
Second, on the basis of native AI and collaboration of core node and SEN layers, we build an innovative technological framework of pervasive multi-level (PML) native AI to further push ${\rm{TK\upmu}}$ connectivity to extreme. Knowledge graph representation is leveraged to generate the feature data set (FDS), which in turn drives the AI algorithm in all nodes and all levels of the system to optimize the quality of service. High-level self-driving and service customization are thus enabled.
Then, a full-spectrum cell-free radio access network (RAN) is designed and deployed within the proposed architecture. A series of emerging 6G technologies including cell-free distributed massive MIMO, photonics-assisted THz-band wireless transmission, and spatiotemporal two-dimensional (2-D) channel coding are applied therein to realize the ${\rm{TK\upmu}}$ KPIs. Finally, a prototype testbed is implemented and experiments are carried out to showcase the advantages of the proposed scheme with respect to the network intelligence, extremely high data rate and SE, etc. The presented promising results can potentially serve as a reference and guideline for the future 6G research and development.


\section{Three-Layer Architecture for 6G}
\label{sec2}

\subsection{Overall Architecture and Design Rationale}

6G networks will exhibit flexible adaptability to diverse vertical applications through new compute plane which coordinates communication and compute in real-time (RT) control manner.
The new compute plane as one of network functions of 6G network will act as the fundamental enabler for agile proximity services, and the decentralized trend will be characterized by the SENs. To this end, we propose an innovative hierarchical architecture design for 6G ${\rm{TK\upmu}}$ extreme connectivity. The rationale behind this architecture design is threefold. First of all, the current ``chimney-style'' mutually independent big data platforms, intelligent platforms, virtualized connection platforms, and storage platforms will be merged into an integrated whole. They will be unified in the form of edge data centers that supplemented by various accelerator and plug-in acceleration cards. Second, a transformation in the morphological characteristics of base stations will be witnessed. More specifically, the accelerated and broadbandized baseband processing function will be gradually integrated into the edge data center. Third, the generation of network functions and application services will demonstrate APP-ization. In particular, the service subscribers will be able to conveniently use the network-side and terminal-side APP platforms to customize the network services and security policies via dynamic configuration of the computing, AI, storage and connectivity resources in the super data center. 

The proposed network architecture can be partitioned into three layers, namely core nodes (CoNs), super edge nodes (SENs) and distributed units. The CoNs constitute the first layer, which is responsible for the non-RT control and management of the entire network.
The second layer (SEN) provides edge agile services including connectivity, computing, data, and AI for the specific service. This is the \emph{key} layer in the 6G network architecture.
The EDUs and the remote radio units (RRUs) are in the third layer of the proposed three-layer architecture. In the following, we elucidate the function and operating mechanism of the SEN layer.

\subsection{Super Edge Nodes}

The SEN layer is logically composed of several separate/converged functions, such as user plane function (UPF) and a portion of the baseband units (BBUs). This layer consolidates edge data acquisition/processing and AI functions to form multiple SENs. Each SEN provides nearby agile services in near-RT according to the task-centric model, which is the basic enabling node for the transformation of 6G wireless networks from \emph{session-centric} to \emph{task-centric} and \emph{decentralized}. The architecture of the SEN network is shown in Fig. \ref{fig2_2}. The specific functions of SEN are implemented by network supermarket (NS), super-RAN intelligent platform (sRIP) and programmable BBU pool (pBBUP).

In terms of sRIP, the architecture of the RAN plays an important role in realizing 6G ${\rm{TK\upmu}}$. The 6G RAN needs to support full spectrum access, including sub-6GHz, mmWave and THz frequency bands. To this end, we need to design a unified implementation of cell-free RAN for different frequency bands. In the proposed cell-free RAN architecture, the EDU module is added to the system to connect with multiple RRUs~\cite{You2021Distributed}. As such, enhanced joint processing ability is yielded and distributed receiver/precoding can be realized at EDU to achieve unlimited scale of cell-free network. On the other hand, the user-centric distributed unit (UCDU) enables decentralized cooperative combination and distribution of user data. The MAC layer and the radio resource control are implemented in centralized control unit (CCU). Therefore, compared with the traditional cell-free massive MIMO, the novel cell-free RAN architecture can achieve better cooperative transmission performance even with a limited-scale RRU~\cite{49}.

Different from 5G, in the 6G network architecture we propose, intelligence is hierarchically and distributedly embedded in edge nodes, from the cloud to each BBU, from L3, L2 to L1. Thus, it is referred to as \emph{native AI}, which provides near-RT/RT intelligent control, owing to its ability to achieve a high level of network autonomy. In the designed RAN architecture, PHY functions are implemented in two separate layers, which reaps the following three benefits. First, it enables the decoupling of high and low frequency systems. Second, it also decouples multi-user data streams, such that the scalability of cell free massive MIMO is enhanced. Third, with EDU-UCDU separation, since the high-PHY processing is user equipment (UE)-oriented, it supports user-centric networks and cloudized pBBUP. Based on the above features, a full-spectrum cell-free RAN architecture was arrived at. On top of 5G, through the utility of virtualized and pooled pBBUs, the proposed cell-free RAN harmoniously integrates with the near-RT/RT resources enabled by native intelligence. As such, flexible and agile allocation of end-to-end full-spectrum air interface resources is accommodated. This in turn fulfills the need for dynamic adaptation of different services to the ${\rm{TK\upmu}}$ air interface access capability, so as to realize the ${\rm{TK\upmu}}$ extreme connectivity towards 6G.

\section{Key Enabling Technologies}
\label{sec3}

\begin{figure*}[!ht]
	\centering
	\subfigure[]{\label{fig3_5}
		\includegraphics[width = 0.8\linewidth]{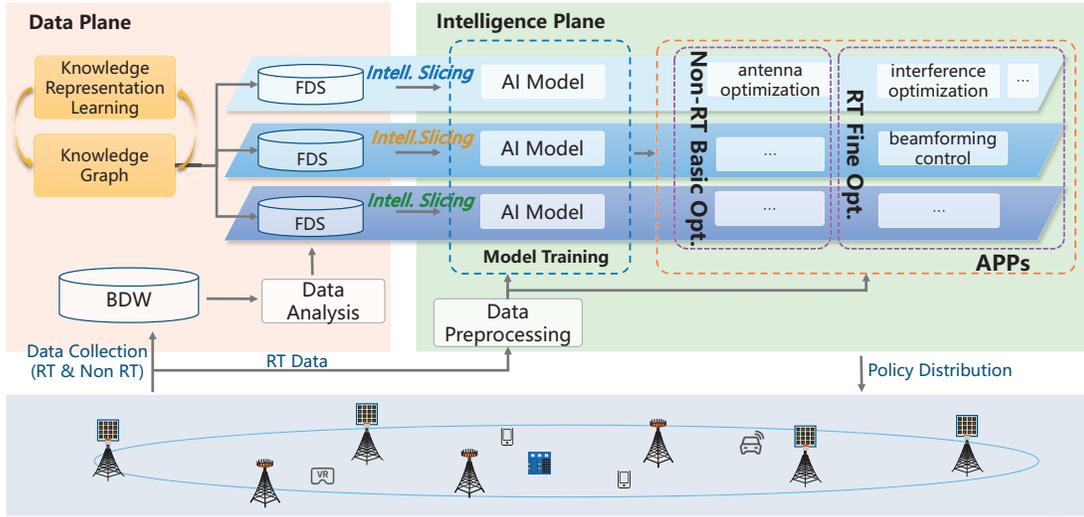}}
	\subfigure[]{\label{fig3_3}
		\includegraphics[width = 0.67\linewidth]{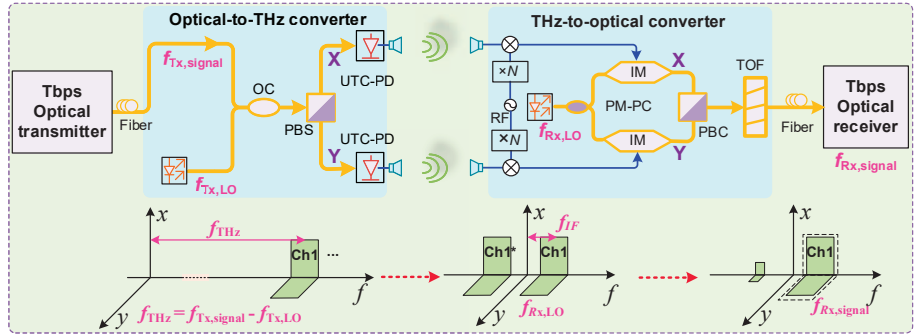}}
	\subfigure[]{\label{fig3_4}
		\includegraphics[width = 0.55\linewidth]{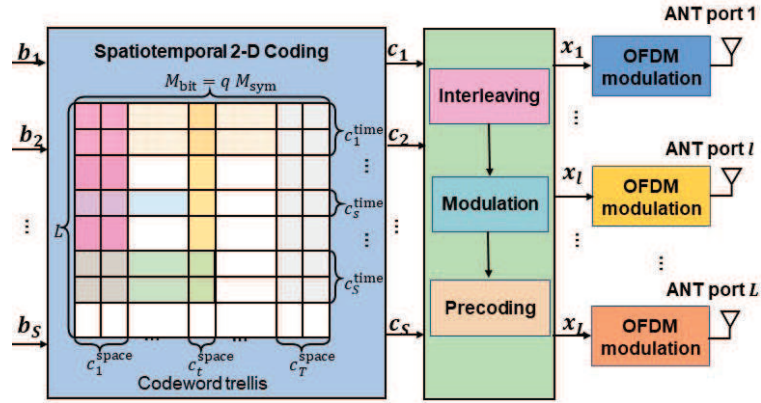}}
	\caption{Key enabling technologies to achieve high data rate and ultra-low latency communications. (a) PML AI architecture. (b) Diagram of a photonics-assisted Tbps fiber-THz-fiber seamless transmission system. (c) Multi-antenna baseband transmitting system with the spatiotemporal 2-D channel coding.}
\end{figure*}

In the previous section, we elaborated the three-layer 6G network hierarchy and explained the potential advantages of the architecture to support the ${\rm{TK\upmu}}$ communications. In this section, we elaborate on the key enabling technologies implemented under this architecture to achieve the expected 6G features and KPIs.
Specifically, a technological framework of PML AI is first proposed to realize task-centric near-RT resource allocation and network automation. Then, a massive MIMO based cell-free architecture is designed as a uniform platform to enable full-spectrum ${\rm{TK\upmu}}$ wireless access. The combination of PML AI and cell-free RAN facilitates agile adaption of ${\rm{TK\upmu}}$ connectivity. Under the new architecture, a few wireless transmission technologies including cell-free massive MIMO, photonics-assisted THz, and spatiotemporal 2-D coding are discussed.

\subsection{Near-RT AI-Empowered Centralized Unit}

At the very inception of 5G networks, incorporating AI natively is beyond the consideration of the designers. As a result, the current 5G systems integrate AI mainly by adopting the add-on and centralized styles, which brings along heavy pressure on transmission bandwidth and formidable challenges in data privacy and protection. In this context, we propose a PML AI technological framework.
By creating innovative data plane and intelligence plane, we infiltratively and pervasively embed the data, computing power and AI in all nodes and all levels of the system. Non-RT, near-RT and RT intelligent optimizations are realized in multiple different and proper levels, so as to shape the true native AI and enable high-level self-management and self-optimization, as well as high-quality servitization of future 6G networks.

Fig. \ref{fig3_5} shows a diagram of the overall architecture of the proposed PML AI, which mainly comprises three parts, i.e., data plane, intelligence plane, and the intelligent slicing connecting the two planes~\cite{11}. The main responsibility of the data plane is to collect data from the network and then analyze the data to dynamically form the FDS as demand, which is to enable true cognitive intelligence that is capable of perceiving and reasoning for diverse applications. However, trillions of data fields and indicators related to heterogeneous software/hardware, functions and protocol stacks are continually generated during the network operation. The network structure, terminal type, terminal behaviour, data service requirements, and system resources of the 6G networks exhibit \emph{high dynamics}, \emph{strong timeliness}, and \emph{mutual coupling}. In aware of the above facts, we propose to use knowledge graph to untangle the complicated connections among the wireless data fields, as well as characterize the types/attributes and the relationships between them. The generated datasets are required to be representative and redundancy-free.

To achieve a high-level end-to-end network autonomy and high-quality guarantee for differential multi-service concurrency requirements, especially agile adaption to varied requirements on the ${\rm{TK\upmu}}$ connectivity ability, the PML AI is expected to allocate multiple dimensional communication/computing/caching resources in task-centric manners via hierarchical intelligence control \cite{35(4.4), Huang2021Patent}. This involves non-RT AI models at the core nodes, near-RT and RT AI models at the SENs and the DUs, respectively. In addition, \emph{statistics of statistics} can help design hierarchical approaches for problem simplification. Therefore, some RT tasks require effective collaboration between non-RT and RT AI models. Furthermore, AI models can impose very high requirements in terms of computing power and communication \cite{36(4.5)}. Distributed deployment of AI models are therefore carefully designed according to the overhead assessment.

\subsection{Full Spectrum Cell-free RAN for 6G TK$\mu$}

In the lower and DU layers, recent evolution in RAN architecture also opens up new opportunities for delivering more agile services and advanced capabilities to end users. Among the candidates for RAN architecture, full spectrum uniform cell-free RAN arguably represents the most appealing one towards 6G ${\rm{TK\upmu}}$. In the following, we introduce a few most promising examples of RAN technologies for 6G ${\rm{TK\upmu}}$.

\subsubsection{Cell-free massive MIMO for Kbps/Hz SE}

Under the cell-free RAN architecture, the spectral efficiency of both uplink and downlink can be improved significantly with collaborative multi-user transmission technology. To achieve ultra-high spectral efficiency and reduce the complexity of implementation, key challenges for the cell-free based transmission include user association, low-complexity transmission methods, and over-the-air TDD reciprocity calibration. In the following, we will introduce the key technologies involved.

In the new architecture, joint transceiver for multiple users and multiple RRUs can be implemented in EDU. Due to the distributed deployment of RRUs, the distances from multiple users to multiple RRUs are different, and the frequency selectivity of the channel is more serious than that of centralized MIMO. Therefore, we derived an optimized precoding vector within multiple resource blocks (RBs) to decrease the complexity of downlink precoding of wideband OFDM systems.

Additionally, in order to reduce the complexity of user-centric collaborative transmission, the association of UCDU with users can be used to achieve dynamic load balancing in the actual deployment.
The acquisition of channel state information is the bottleneck of cell-free system. Due to the large number of users and limited pilot overhead, pilot reuse needs to be studied. For the cell free RAN architecture, the RRUs managed by EDU send calibration signals to each other and exchange the received calibration signals. The method in literature \cite{You2021Distributed} can be used to obtain the reciprocity calibration coefficients of RRUs.

\subsubsection{Photonics-assisted THz wireless access for Tbps Rate}

The existing communication technology is incompetent to meet the explosive data rate requirements of ultra-high bandwidth communication networks. This motivated the exploration of alternative radio-frequency spectrum including the THz-band ($0.3-10\;{\rm{THz}}$). Therefore, we propose a photonics-assisted Tbps fiber-THz-fiber seamless integration system, which is shown in Fig.~\ref{fig3_3}. The prototype system mainly contains three key components: Tbps optical transceivers, optical-to-THz (O/T) module, and THz-to-optical (T/O) module. The baseband polarization-division-multiplexing (PDM) optical signal ($f_{\text{Tx,signal}}$) is generated from optical transmitters and transmitted over optical fiber. At the optical-to-THz (O/T) conversion front-end, optical signals are heterodyne beat with the local oscillator ($f_{\text{Tx,LO}}$) via ultra-fast uni-travelling-carrier photodiode (UTC-PD) to generated PDM THz-wave signals ($f_{\text{THz}} = f_{\text{Tx, signal}} - f_{\text{Tx, LO}}$). After $2 \times 2$ MIMO wireless transmission, at the THz-to-optical (T/O) conversion front-end, analog down conversion is first operated to obtain intermediate frequency (IF) signals ($f_{\text{IF}}$). Then, the IF signals are converted into PDM double-sideband optical signals ($f_{\text{Rx,LO}}$) via two parallel intensity-modulators (IMs). Finally, one sideband filtering from an optical filter as baseband optical signal ($f_{\text{Rx,signal}}$) is delivered over optical fiber and fed into the optical receivers. At the Tbps transceivers, high-order modulation formats and advanced digital signal processing (DSP) techniques are utilized to reach Tbps transmission capacity.

This seamless integrated fiber-THz-fiber scheme has many potential application scenarios. For instance, THz wireless transceivers can be used to replace optical fibers or cables to achieve high-speed wireless backhaul transmission among base stations in areas where optical fibers cannot be deployed. Besides, this conceptual scheme can potentially provide emergency communication services to replace the interrupted large-capacity long-distance fiber link during the natural disasters including hurricane, earthquakes and flood. THz wireless communication between inter-rack and intra-rack is considered to be used in data centers due to its ultra-high communication rate to save data center space costs and cable maintenance costs. The proposed architecture can meet the demands of future THz seamless communication with low power consumption, low cost and miniaturization.

\subsubsection{Spatiotemporal 2-D channel coding in massive MIMO for $\mu$s-level latency}

Aside from the data rate and SE, latency is another important KPI in the 6G networks. The 6G network is envisioned to facilitate more URLLC applications with more rigorous latency constraint, which cannot be supported by 5G networks. Traditional channel coding schemes are commonly applied in the time domain and the decoding latency is high for long packets (such as eMBB services), while the reliability is degraded for short packets in the URLLC applications. As a matter of fact, very limited room is left for channel coding design in the time domain, since 6G systems require $\rm{\upmu s}$-level transmission with high reliability. On the other hand, as hundreds of antennas are employed in the cell-free massive MIMO system, the extra degree of freedom in the space domain can be fully utilized, and this motivates us to propose the spatiotemporal 2-D channel coding scheme. The channel coding is based on the codeword trellis shown in Fig. \ref{fig3_4}. Specifically, the bit-stream are encoded in the time domain, and then a space-domain encoding is performed based on the time domain codes. 
At the receiver side, log-likelihood ratios (LLRs) of different space-domain decodings are performed in parallel, and then the time-domain decodings are proceeded with the outputs of decoded information bits. By carefully designing the spatiotemporal 2-D coding scheme, extremely low-latency transmissions at $\rm{\upmu s}$-level can be achieved in the 6G networks with cell-free massive MIMO systems~\cite{50}.

\begin{figure*}[!htb]
	\centering{\includegraphics[width = 0.82\linewidth]{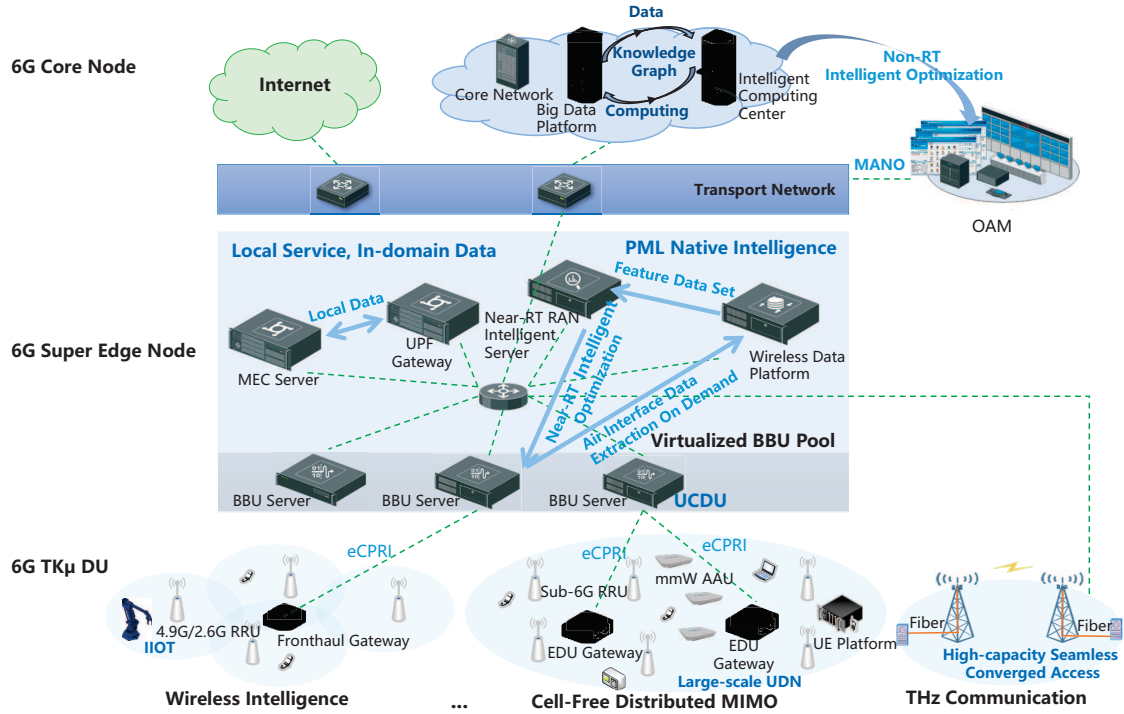}}
	\caption{Implementation diagram of the proposed three-layer 6G network.
		\label{fig4_1}}
\end{figure*}

\section{Implementation and Experiment Evaluations of Open Testbed}
\label{sec4}

Taking the proposed 6G wireless network architecture characterized by SEN as the blueprint, we built the sub-systems of 6G CoN, SEN and DU, on the basis of an open 5G platform. As illustrated in Fig.~\ref{fig4_1}, by assembling these components together, we established the world's first task-centric, integrated, intelligent and open 6G testbed that supports ${\rm{TK\upmu}}$ extreme connectivity.
The testbed supports experimental verification and performance evaluation of the innovative technologies for PML AI and ${\rm{TK\upmu}}$ full spectrum cell-free distributed massive MIMO. The implementation of the full spectrum cell-free RAN of the testbed is based on the open interface. It supports sub-6GHz and mmWave systems and THz communications, as well as ultra-low-latency and ultra-high-reliability transmission. For cell-free massive MIMO in sub-6GHz and mmWave bands, taking $100\;{\rm{MHz}}$ bandwidth as a baseline, a single EDU can support 16T16R and be further extended to 64T64T. In realizing the PML native AI network intelligence, the SEN is capable of performing RT complete acquisition of wireless network air-interface data, based on which the knowledge graph of network data is established in the CoN in a non-RT way. To support the near-RT intelligent optimization for varying applications, the knowledge-graph-based FDS is generated on-demand in the wireless data platform in the SEN, and further used in the near-RT RAN intelligent server to implement AI training/reference and the resultant dispatch in the BBUs, where $20\;{\rm{ms}}$ control granularity can be achieved.

\begin{figure*}[!htb]
	\centering
	\subfigure[]{\label{Fig_4_MIMO}
		\includegraphics[width = 0.9\linewidth]{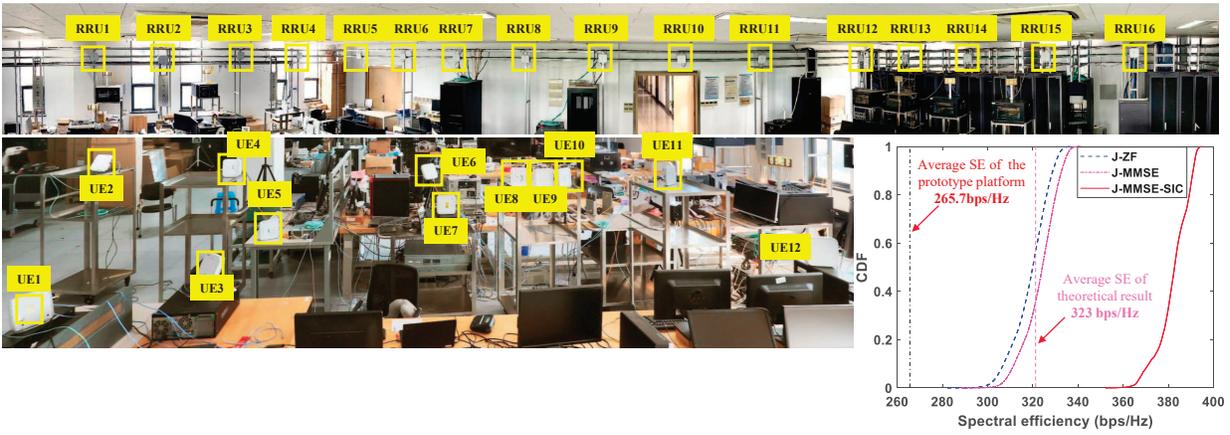}}
	\subfigure[]{\label{Terahertz_2}
		\includegraphics[width = 0.9\linewidth]{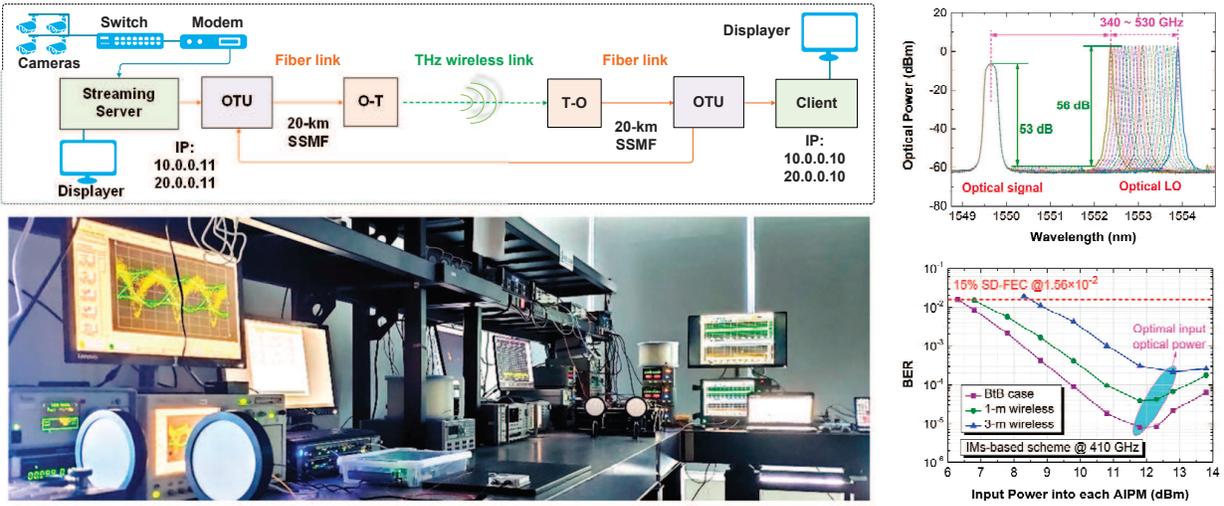}}
	\subfigure[]{\label{fig4_ICIC}
		\includegraphics[width = 0.9\linewidth]{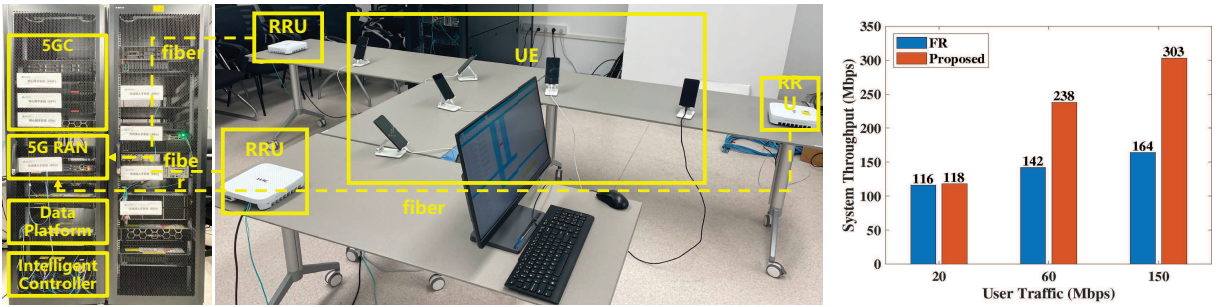}}
	\caption{Experiment configurations and results. (a) Test of cell-free massive MIMO technologies. (b) Test of photonics-assisted THz-Band wireless access (courtesy of Zhang et al.\cite{zhang2022real}). (c) Implementation and performance evaluation of the intelligent ICIC case.}
\end{figure*}

\subsection{Experiment of Cell-Free Massive MIMO Based on COTS-RRU}

We develop a cell-free massive MIMO prototype platform shown in Fig. \ref{Fig_4_MIMO} as a demo system of the 6G DU to verify the cell-free distributed MIMO technologies in the 6G testbed.
Specifically, the 5G commercial-off-the-shelf (COTS) RRUs are used for both base-station and UEs, and the interface between RRU and baseband unit is compatible with eCPRI~\cite{Cao2022Experimental}.
The hardware specifications of the UE and BS are identical.
The system operates in the $4.9$ GHz band, and the RRU of each BS/UE is equipped with $4$ antennas.
All the RRUs for the BS share the same reference clock from the BBU.
Each RRU uses an independent local oscillator.

The prototype system can serve $12$ UEs and realize parallel transmission of $48$ data streams. Each UE use $64$QAM and $0.89$ code rate of LDPC, with considering the overhead of pilot and cyclic prefix, the real time average spectral efficiency of the system reaches $209$ bps/Hz. We perform the offline spectral efficiency analysis on the channels collected from the prototype system. The result shows that when using joint processing linear minimum mean square error (LMMSE) detection, theoretically the average spectral efficiency can achieve $323$ bps/Hz. For the real time spectral efficiency, if we neglect the overhead, the average spectral efficiency of the system is $265.7$ bps/Hz, which can achieve the $82\%$ of the theoretical result. Compared with 5G small cell system with the same deployment, the cell-free massive MIMO system can get $10\times$ performance improvement.

\subsection{Experiment of Photonics-Assisted THz-Band Wireless Access}

To further confirm high data-rate transmission capability at THz-band and the associated access technologies in the 6G testbed, we build an RT single-channel transparent fiber-THz-fiber $2 \times 2$ MIMO seamless transmission system with a record net rate of single-channel $103.125$ Gbps and dual-channel $206.25$ Gbps at WR2.2 band ($330 – 500$ GHz) using the commercial digital coherent optics (DCO) modules~\cite{zhang2022real}. Considering low power consumption and miniaturization, O/T conversion at the transmitter is based on photomixing in a UTC-PD, and T/O conversion at the receiver is realized using hybrid optoelectronic down-conversion.
As shown in Fig. \ref{Terahertz_2}, the $125.516$ Gbps line data rate adopting $31.379$ GBaud dual polarization quadrature-phase-shift-keying (DP-QPSK) modulation is successfully transmitted over a wireless distance of $3$ m, two spans of $20$ km fiber with $15\%$ soft-decision forward-error-correction (SD-FEC) of $1.56 \times 10^{-2}$.

We also develop a $100/200$ GbE streaming service platform to test the stability of the transmission system by playing RT movie and live surveillance video. This is the first reported demonstration to realize $> 100$ Gbps RT transparent fiber-THz-fiber link transmission at beyond $350$ GHz band. This proof-of-concept study is certainly a concrete and inspiring step towards practical transparent fiber-THz-fiber link for future ${\rm{TK\upmu}}$-standard 6G mobile communication system.

\subsection{Experiment of Intelligent Network Optimization}
		
As the third functional test, we validate the PML native AI in the testbed by evaluating the performance of intelligent inter-cell interference coordination (ICIC). The proposed PML architecture allows finer-grained parameter control in terms of tunable parameter numbers and time granularity. In particular, as shown in Fig. \ref{fig4_ICIC}, the proposed intelligent ICIC scheme trains and maintains a reinforcement learning-based model in the non-RT basic optimization layer. By referring to the FDS distilled from the data plane, it adaptively adjusts the parameters in the resource allocation module deployed in the RT fine optimization layer, with a time granularity over hundreds of milliseconds. In the meantime, a resource allocation module based on the conflict graph model runs in the RT fine optimization layer, which operates in the order of hundreds of milliseconds or below. This module first intelligently identifies interferences among the UEs and the neighboring cells based on the interference FDS, i.e., reference signal received power (RSRP). Then, the physical resource block (PRB) resource sets are preconfigured and the cells schedule UEs in real-time accordingly. The non-RT basic optimization layer and RT fine optimization layer constitute a model-data dual-driven near-RT intelligent ICIC solution, which supports response in the order of hundreds of milliseconds. 

In the testbed experiments, a CoN and a RAN intelligent server act as the non-RT basic optimization layer and RT fine optimization layer, respectively. Additionally, three RRUs and six UEs are deployed in the trials, as shown in Fig. \ref{fig4_ICIC}. In the experiments, for UEs in each cell, we place one close to the RRU and another far from the RRU. Ipef software is used to generate constant traffic for UEs, and the traffic of each UE varies from $20$ Mbps to $150$ Mbps. For comparison, the full reuse (FR) scheme, i.e., each cell can use the entire bandwidth, is chosen as the baseline. The results reveal that the proposed ICIC method can significantly improve the system throughput, especially in the case of heavy user traffic. Specifically, as the user traffic rises from $20$ Mbps to $150$ Mbps, the system throughput achieved by FR slowly climbs from $116$ Mbps to $164$ Mbps. By contrast, a sharp increase from $118$ Mbps to $303$ Mbps is achieved by the proposed ICIC scheme. As indicated in Fig. \ref{fig4_ICIC}, the system throughput is increased by $2\%$, $68\%$, and $85\%$ respectively. Note that inter-cell interference (ICI) deteriorates in the case of heavy user traffic and, therefore, it dominates the throughput. This implies that the proposed ICIC method can remarkably suppress the ICI.

\section{Conclusion}
\label{sec5}

In this article, we presented systematic architecture, technological framework and testbed implementation towards the visions and expectations of 6G ${\rm{TK\upmu}}$ extreme connectivity.
The overall architecture was totally redesigned to fully integrate the multi-dimensional resources and functions, such that a pervasive, converged and intelligent information infrastructure can be established.
The SEN, as the core element therein, facilitates task-centric and decentralized operation paradigm and, by virtue of which, high-quality service customization and high-degree network autonomy can be realized. In addition, we developed and incorporated the advanced PHY technologies such as cell-free distributed massive MIMO, photonics-assisted THz-band access and spatiotemporal 2-D channel coding in the proposed three-laryer hierarchy. As such, a uniform full-spectrum radio access can potentially be achieved for ${\rm{TK\upmu}}$-level KPIs of full-service and full-scenario applications. Extensive experiments were carried out to test the individual sub-systems designed for ${\rm{TK\upmu}}$ and native intelligence. Some possible directions of our future work include update of the three-layer intelligent architecture, development of more emerging technologies and methodologies, accommodation of other critical KPIs such as $90\%$ coverage via territorial-satellite convergence, etc.

\section*{Acknowledgment}

This work was supported in part by the National Key Research and Development Program under grants 2020YFB1806600, 2020YFB1807200, 2021YFB2900300, 2018YFE0205900, and the Major Key Project of Peng Cheng Laboratory under Grant PCL 2021A01-2.

\balance
\bibliographystyle{IEEEtran}
\bibliography{TKmu_WCM}

\begin{thebibliography}{10}
\providecommand{\url}[1]{#1}
\csname url@samestyle\endcsname
\providecommand{\newblock}{\relax}
\providecommand{\bibinfo}[2]{#2}
\providecommand{\BIBentrySTDinterwordspacing}{\spaceskip=0pt\relax}
\providecommand{\BIBentryALTinterwordstretchfactor}{4}
\providecommand{\BIBentryALTinterwordspacing}{\spaceskip=\fontdimen2\font plus
\BIBentryALTinterwordstretchfactor\fontdimen3\font minus
  \fontdimen4\font\relax}
\providecommand{\BIBforeignlanguage}[2]{{%
\expandafter\ifx\csname l@#1\endcsname\relax
\typeout{** WARNING: IEEEtran.bst: No hyphenation pattern has been}%
\typeout{** loaded for the language `#1'. Using the pattern for}%
\typeout{** the default language instead.}%
\else
\language=\csname l@#1\endcsname
\fi
#2}}
\providecommand{\BIBdecl}{\relax}
\BIBdecl

\bibitem{NSF2021}
\BIBentryALTinterwordspacing
{Next G Alliance white paper}. (2022) {Next G Alliance} report: {Roadmap} to
  {6G}. [Online]. Available:
  \url{https://nextgalliance.org/wp-content/uploads/2022/02/NextGA-Roadmap.pdf}
\BIBentrySTDinterwordspacing

\bibitem{Hexa2020}
\BIBentryALTinterwordspacing
{Hexa-X}. (2022) {Hexa-X} vision on {6G} and research challenges. [Online].
  Available: \url{https://hexa-x.eu/vision/}
\BIBentrySTDinterwordspacing

\bibitem{Samsung2020}
\BIBentryALTinterwordspacing
{Samsung 6G vision white paper}. (2020) {6G: The next hyper-connected
  experience for all}. [Online]. Available:
  \url{https://cdn.codeground.org/nsr/downloads/researchareas/20201201_6G_Vision_web.pdf}
\BIBentrySTDinterwordspacing

\bibitem{XR}
M.~Giordani, M.~Polese, M.~Mezzavilla, S.~Rangan, and M.~Zorzi, ``Toward {6G}
  networks: {Use} cases and technologies,'' \emph{IEEE Communications
  Magazine}, vol.~58, no.~3, pp. 55--61, 2020.

\bibitem{5}
W.~Saad, M.~Bennis, and M.~Chen, ``A vision of {6G} wireless systems:
  {Applications}, trends, technologies, and open research problems,''
  \emph{IEEE Network}, vol.~34, no.~3, pp. 134--142, 2019.

\bibitem{tariq2020speculative}
F.~Tariq, M.~R. Khandaker, K.-K. Wong, M.~A. Imran, M.~Bennis, and M.~Debbah,
  ``A speculative study on {6G},'' \emph{IEEE Wireless Communications},
  vol.~27, no.~4, pp. 118--125, 2020.

\bibitem{11}
Y.~Huang, S.~Liu, C.~Zhang, X.~You, and H.~Wu, ``True-data testbed for {5G/B5G}
  intelligent network,'' \emph{Intelligent and Converged Networks}, vol.~2,
  no.~2, pp. 133--149, 2021.

\bibitem{You2021Distributed}
X.~You, D.~Wang, and J.~Wang, \emph{Distributed {MIMO} and Cell-free Mobile
  Communication}.\hskip 1em plus 0.5em minus 0.4em\relax Springer, 2021.

\bibitem{49}
D.~Wang, ``Wireless transmission techniques of cell-free massive {MIMO} for
  {6G} mobile communications,'' \emph{Mobile Communications}, vol.~4, no.~7,
  pp. 10--15, 2021.

\bibitem{35(4.4)}
S.~Liu, C.~Zheng, Y.~Huang, and T.~Q. Quek, ``{Distributed reinforcement
  learning for privacy-preserving dynamic edge caching},'' \emph{IEEE Journal
  on Selected Areas in Communications}, vol.~40, no.~3, pp. 749--760, 2022.

\bibitem{Huang2021Patent}
Y.~Huang, J.~Li, C.~Zhang, and X.~You, ``Network slice dynamic configuration
  method, device, electronic device and storage medium,'' Chinese Patent, No.
  202110336480.9, 2021.

\bibitem{36(4.5)}
C.~Xu, S.~Liu, Z.~Yang, Y.~Huang, and K.-K. Wong, ``Learning rate optimization
  for federated learning exploiting over-the-air computation,'' \emph{IEEE
  Journal on Selected Areas in Communications}, vol.~39, no.~12, pp.
  3742--3756, 2021.

\bibitem{50}
X.~You, C.~Zhang, B.~Sheng, Y.~Huang, C.~Ji, Y.~Shen, W.~Zhou, and J.~Liu,
  ``Spatiotemporal {2-D} channel coding for very low latency reliable {MIMO}
  transmission,'' \emph{arXiv preprint arXiv:2201.03166}, pp. 1--7, 2022.

\bibitem{zhang2022real}
J.~Zhang, M.~Zhu, B.~Hua, M.~Lei, Y.~Cai, Y.~Zou, W.~Tong, J.~Ding, L.~Tian,
  L.~Ma \emph{et~al.}, ``Real-time demonstration of 100 gbe thz-wireless and
  fiber seamless integration networks,'' \emph{Journal of Lightwave
  Technology}, 2022.

\bibitem{Cao2022Experimental}
Y.~Cao, P.~Wang, K.~Zheng, X.~Liang, D.~Liu, M.~Lou, J.~Jin, Q.~Wang, D.~Wang,
  Y.~Huang \emph{et~al.}, ``Experimental performance evaluation of cell-free
  massive {MIMO} systems using {COTS RRU} with {OTA} reciprocity calibration
  and phase synchronization,'' \emph{arXiv preprint arXiv:2208.14048}, 2022.

\end{thebibliography}

\end{document}